# Assessment of classification techniques on predicting success or failure of software reusability

Nahid Hajizadeh, Manijeh Keshtgari , Marzieh Ahmadzadeh
Department of computer engineering and IT
Shiraz University of technology
Shiraz, Iran

*Abstract*— Software reusability means use of reusable modules when development of the software takes place so that it not needs to generate code from scratch. This capability not only reduces software development time and cost of development but also improves software productivity. To predict whether a software product under development will be used as a reusable module or not in the future, will assist software project managers to ensure that the product is reusable, otherwise modify the weak points. In this paper, to predict success or failure of software reusability, 42 classification algorithms on a specific Software reuse data set (known as PROMISE Software Engineering Repository data set [1]) are applied. By comparing 8 conventional evaluation metrics obtain by each of 42 implementation of the classification algorithms, results showed that "SGD or SMO" and "Multi Class Classifier Updateable" algorithms has a better performance in comparison with other algorithms.

*Keywords-* Data mining, classification, Software reuse, Assesment.

## I. INTRODUCTION

Software reusability, besides reducing the time and cost of software development processes increases productivity. Software reusability not only stands for using part of one system in other system, but also using whole part of other systems is welcomed [2], [3]. Ramamortty et al [4] noted that in addition to reusing source code, you will find that other software projects assets such as requirements, design, test cases and documents could be reused. Software project managers are extremely interested that during developing their software products and before its offering to the market to ensure that it is reusable. Since the main application of classification algorithms are to predict, we also used classification algorithms for prediction of success or failure of software reusability prior to marketing. Classification algorithms are a main branch in data mining science. Data mining is the main step in Knowledge Discovery in Databases (KDD) process. Though KDD is used synonymously to represent data mining, both these are actually different. Some preprocessing steps before data mining are to be done to transform the raw data as useful knowledge. Steps of knowledge discovery are selection of data set, data preprocessing, data transformation, data mining and knowledge extraction, respectively (Figure 1 [5]).

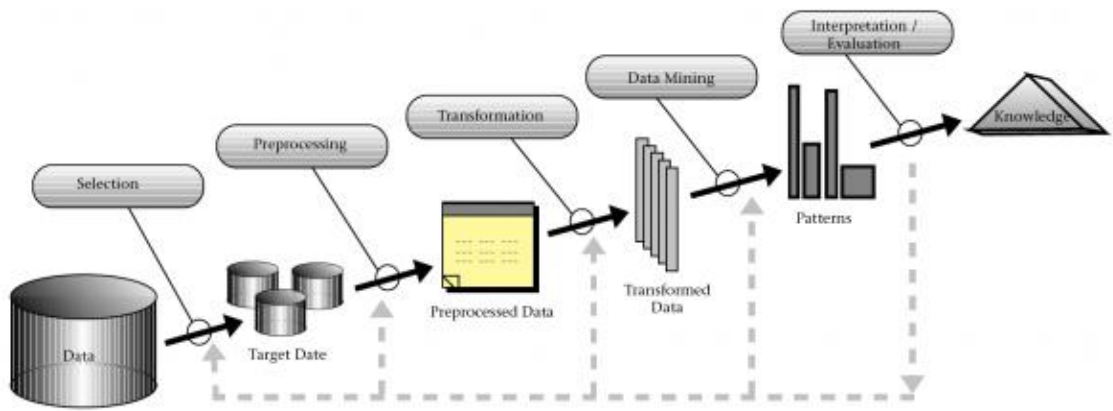

Figure 1.   DataMining process

## II. RELATED WORK

Until now, many researches on Software Reusability based on data mining have been done by various authors. Morisio et.al [6] has introduced a number of the key factors in predicting software Reusability. Clustering methods for predicting reusability of object oriented software systems has been done by Shri et.al [7]. An artificial neural network method has been suggested by G.Boetticher et.al [8] to rank the reusability of software products. In [9] Basili et.al indicated that most of the metrics proposed by Chidamber and Kemerer in [10] are useful for predicting the fault-proneness of classes during the design phase of Object Oriented systems [11]. Sonia et.al [12] has proposed a framework for evaluating reusability of procedure oriented system using metrics based approach.

## III. METHODOLOGY

The data set that employed in this paper is provided by Morisio et al [6] and is one of the PROMISE Software Engineering Repository data set [1]. This data set represents an interesting SE problem: how to make sense of software development when very little data (only 24 instances) is available. This Data set contains 27 attributes and a class variable.

## IV. DATA PREPROCESSING PHASE

According to missing values in data set, doing preprocessing is not inevitable.

In this paper for handling of missing attribute values, "Mode" technique has been used in such a manner that values of missing attributes are set due to majority values of same attribute.

For example the value of 'Development Approach' for instance 24 is missed (Table 1)

The most probable value for attribute 'Development Approach' is "OO", so value "OO" is selected for this missing value (Table 2)

TABLE 1. 'DEVELOPMENT APPROACH' FOR INSTANCE 24 (BEFORE PREPROCESS)

| 1 | 2 | 3 | 4 | 5 | 6 | 7 | 8 | 9 | 10 | 11 | 12 | 13 | 14 | 15 | 16 | 17 | 18 | 19 | 20 | 21 | 22 | 23 | 24 |
|---|---|---|---|---|---|---|---|---|----|----|----|----|----|----|----|----|----|----|----|----|----|----|----|
| OO | OO | OO | OO | OO | OO | OO | OO | OO | proc | proc | proc | proc | OO | proc | proc | proc | OO | OO | OO | OO | OO | proc | **not_available** |

TABLE 2. 'DEVELOPMENT APPROACH' FOR INSTANCE 24 (AFTER PREPROCESS)

| 1 | 2 | 3 | 4 | 5 | 6 | 7 | 8 | 9 | 10 | 11 | 12 | 13 | 14 | 15 | 16 | 17 | 18 | 19 | 20 | 21 | 22 | 23 | 24 |
|---|---|---|---|---|---|---|---|---|----|----|----|----|----|----|----|----|----|----|----|----|----|----|----|
| OO | OO | OO | OO | OO | OO | OO | OO | OO | proc | proc | proc | proc | OO | proc | proc | proc | OO | OO | OO | OO | OO | proc | **OO** |

## V. APPLYING DATA MINING

WEKA toolkit [13], version 3.7.11, with 10-fold cross validation is employed for making models by 42 classification algorithms from 7 main groups. This 42 classification algorithms is listed in Table 3.

TABLE 3. CLASSIFICATION ALGORITHMS

| Group Name | Functions |
|---|---|
| Bayes | BayesNet, NaiveBayes, NaiveBayesMultinomialText, NaiveBayesUpdateable |
| Functions | Logistic, MultilayerPerceptron, SGD, SGDText, SimpleLogistic, SMO, VotedPerceptron |
| Trees | DecisionStump, HoeffdingTree, J48, LMT, RandomForest, RandomTree, REPTree |
| Lazy | IB1, KStar, LWL |
| Rules | DecisionTable, JRip, OneR, PART, ZeroR |
| Meta | AdaBoostM1, AttributeSelectedClassifier, Bagging, ClassificationViaRegression, CVParameterSelection, FilteredClassifier, LogitBoost, MultiClassClassifier, MultiClassClassifierUpdateable, MultiScheme, RandomCommittee, RandomizableFilteredClassifier, RandomSubSpace, Stacking, Vote |
| Misc | InputMappedClassifier |

## VI. EMPIRICAL DATA ANALYSIS

To assessment models made by those 42 different classifiers, Kiviat diagram [14] is applied, so normalization of metric values to scale data to fall within a specified range is necessary. In this work, "normalization by decimal scaling" [15] is used that is shown in equation (1).

$$V' = \frac{V}{10^j} \quad (1)$$

Where j is the smallest integer such that $Max(V') < 1$

Which Metrics applied to assess the performance of these models is as follow:

Incorrectly Classified Instances (ICI) is the number of instances that are incorrectly classified.

Accuracy is the measure to determine the accuracy of a classifier. This measure indicates that what percentage of the total test set records correctly classified. Equation (2) shows the calculation of accuracy.

$$Accuracy = \frac{TN+TP}{TN+FN+TP+FP} \quad (2)$$

Precision for a class is the number of true positives (TP) (i.e. the number of items correctly labeled as belonging to the positive class) divided by the total number of elements labeled as belonging to the positive class [16]. The precision equation is indicated in equation (3) [11].

$$Precision = \frac{TP}{TP+FP} \quad (3)$$

Recall in this perspective is defined as the number of true positives divided by the total number of elements that actually belong to the positive class (i.e. the sum of true positives and false negatives (FN), which are items which were not labeled as belonging to the positive class but should have been). The recall can be calculated as equation (4) [11].

$$Recall = \frac{TP}{TP+FN} \quad (4)$$

To calculate the Cost, confusion matrix [17] element must be multiplied to corresponding entries in the cost matrix and results be added together.

Gain ratio is used to determine the goodness of a split [18].

Mean absolute error (MAE) is the average of the difference between predicted and actual value in all test cases; it is the average prediction error. The formula for calculating MAE is given in equation (5) [11].

$$MAE = \frac{(p_1+a_1)^2 + (p_2+a_2)^2 + ... + (p_n+a_n)^2}{n} \quad (5)$$

The final metric is Relative Absolute Error (RAE) that is used to show Error rate in relative mode. Actual target values and Predicted target values are two parameters to calculate RAE as mentioned in equation (6).

$Actual\ target : a_i, Peredicted\ target : p_i$

$$RAE = \frac{|p_1-a_1|+...+|p_n-a_n|}{|\bar{a}-a_1|+...+|\bar{a}-a_n|} \quad (6)$$

VII. RESULTS

42 classifier algorithms of 7 groups were analyzed and the best classifier of each group is shown in Table 4.

TABLE 4. BEST CLASSIFIER FROM EACH 7 CATEGORIES

| | Accuracy | Cost | Recall | ICI | Gain | MAE | Precision | RAE |
|---|---|---|---|---|---|---|---|---|
| **Bayes (NaiveBayes)** | 95.8% | 1 | 0.958 | 1 | 10 | 0.0608 | 0.961 | 12.795% |
| **Functions (SGD or SMO)** | 95.8% | 1 | 0.958 | 1 | 10 | 0.0417 | 0.961 | 8.773% |
| **Trees (HoeffdingTree)** | 95.8% | 1 | 0.958 | 1 | 10 | 0.0608 | 0.961 | 12.795% |
| **Lazy (LWL)** | 95.8% | 1 | 0.958 | 1 | 10 | 0.0503 | 0.961 | 10.592% |
| **Rules (JRip,PART)** | 95.8% | 1 | 0.958 | 1 | 10 | 0.0794 | 0.961 | 16.71% |
| **Meta (MultiClassClassifierUpdateable)** | 95.8% | 1 | 0.958 | 1 | 10 | 0.0417 | 0.961 | 8.773% |
| **Misc (InputMappedClassifier)** | 62.5% | 9 | 0.625 | 9 | 0 | 0.474 | 0.391 | 100% |

Kiviat diagram of these 7 best classifiers is illustrated in Figure2.

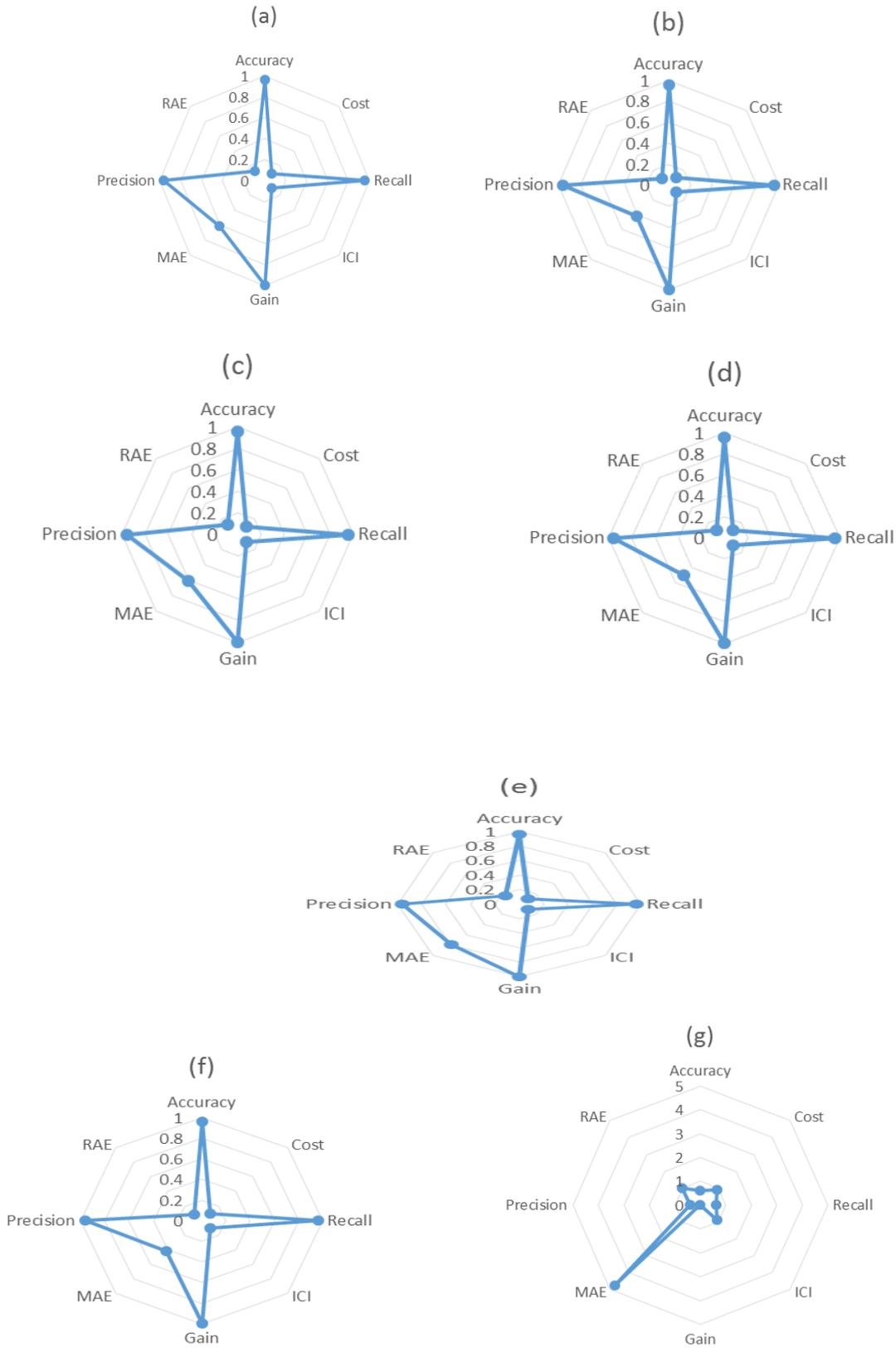

Figure 2.   Kiviat diagram of 7 best classifiers: (a) NaiveBayes, (b) SGD, (c) HoeffdingTree, (d) LWL, (e) JRip, (f) MultiClassClassifierUpdateable, (g) InputMappedClassifier.

Exception of Input Mapped Classifier that in comparison with other types of classifiers has extremely weak performance, , other 6 classifiers have similar values in most metrics and have difference only in Mean Absolute Error(MAE) and Relative absolute error(RAE).

In order to show more clearly the differences and choose the best classifier, Figure 3 which contains a comparison of that 7 classifiers based on two metrics MAE and ARE is plotted.

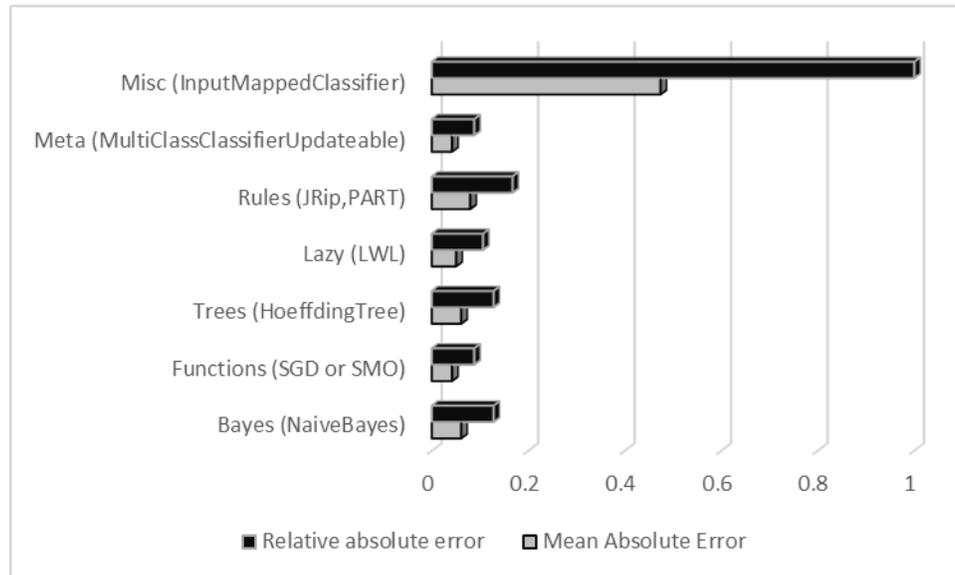

Figure 3. Comparison between 7 best classifiers based on MAE and ARE metrics.

"*SGD or SMO*" and "*Multi Class Classifier Updateable*" have the sample and best performance among others (according to lower value in errors).

## VIII. CONCLUSION

In comparison was made between the different models of classifiers, in general, "*Functions*" and "*Meta*" categories showed better performance than other groups and among these categories, "*SGD or SMO*" and "*Multi Class Classifier Updateable*" classifiers are the most appropriate classifiers for predicting success of failure of software reusability.

## IX. FUTURE WORK

Data set used in this paper does not consider recent software development technologies such as Component base development and Agent base development, while these techniques play an important role in increasing the Software reusability. As future work, to generalize our results, we need to analyze other datasets and representative case studies.